\documentclass[a4paper,
               biblatex,     % biblatex is used
               nospread,     % flushend option: do not fill with whitespace to balance columns
               ]{jacow}

\usepackage[utf8]{inputenc}
\usepackage[USenglish]{babel}

\newcommand{\bslabel}[1]{{\bfseries #1.}}

\usepackage{float}
\usepackage{hyperref}

\PassOptionsToPackage{breaklinks}{hyperref}
\RequirePackage{hyperref}
\definecolor{darkblue}{rgb}{0, 0, 0.5}
\hypersetup{colorlinks=true, citecolor=darkblue, linkcolor=darkblue, urlcolor=darkblue}

\usepackage{booktabs} % For formal tables
\usepackage{graphicx}
\DeclareGraphicsExtensions{.pdf,.ai,.jpg,.png}
\setkeys{Gin}{pagebox=artbox}% Alternative for \pdfpagebox 5 in preceding line.
\graphicspath{{./figures/}}

\begin{document}

\title{\mbox{\kern-1.1em WARC-DL: Scalable Web Archive Processing for Deep Learning}\hspace*{-16pt}}

\author{Niklas Deckers \qquad Martin Potthast\\ Leipzig University}

\maketitle

\begin{NoHyper}
\renewcommand{\thefootnote}{}
\footnotetext{Contact: $<$firstname$>$.$<$lastname$>$@uni-leipzig.de\\This work was supported by the German Federal Ministry of Education and Research (BMBF, 01IS18026B) by funding the competence center for Big Data and AI ``ScaDS.AI'' Dresden/Leipzig.}
\renewcommand{\thefootnote}{\number\value{footnote}}
\end{NoHyper}

\begin{figure}[t]
\vspace{-1.5ex}
\setlength{\hfuzz}{1.1\columnwidth}
\begin{minipage}{\textwidth}
\begin{figure}[H]
\centering
\includegraphics*[width=\textwidth]{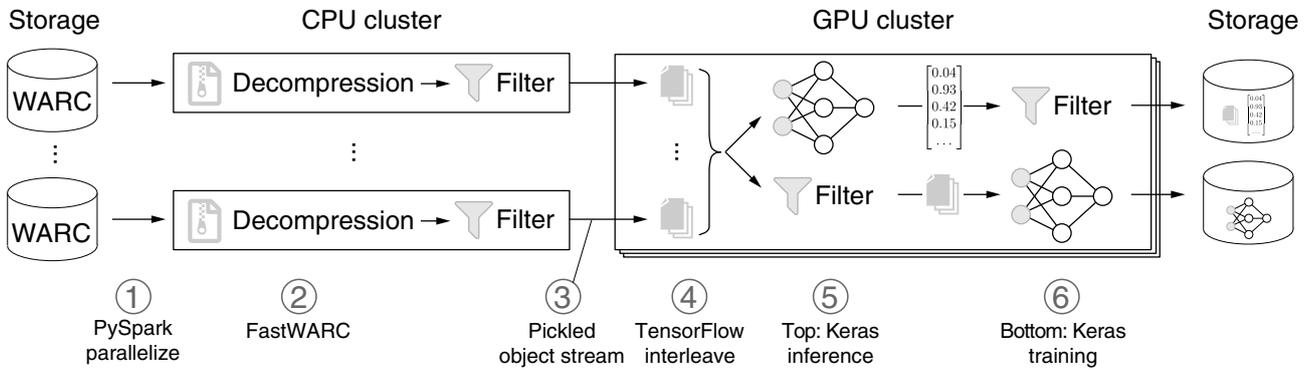}
\caption{The six steps of the WARC-DL web archive processing pipeline: Filters and the Keras training are customizable.}
\label{fig:pipeline}
\end{figure}
\end{minipage}
\vspace{-2.5ex}
\end{figure}

\bslabel{Introduction}
Web archives have grown to petabytes. In addition to providing invaluable background knowledge on many social and cultural developments over the last 30~years, they also provide vast amounts of training data for machine learning. To benefit from recent developments in Deep Learning, the use of web archives requires a scalable solution for their processing that supports inference with and training of neural networks. To date, there is no publicly available library for processing web archives in this way, and some existing applications use workarounds~\cite{yang:2019}. This paper presents WARC-DL,%
\footnote{\url{https://github.com/chatnoir-eu/chatnoir-warc-dl}}
a deep learning-enabled pipeline for web archive processing that scales to petabytes.

\bslabel{Technical Setting}
In many ``traditional'' data centers, mass storage, CPU-bound processing, and GPU-bound processing are separate clusters. Since the storage capacity of the latter two does not usually match that of the former, large web archives are traditionally processed in batches. However, batch processing of web archives on GPU clusters is wasteful: since web archive data is raw data, it must be preprocessed before being passed to a~GPU. This preprocessing is usually CPU-bound and highly parallelizable, so using a CPU~cluster for this is desirable before streaming the preprocessed data to the GPU~cluster.
In addition, the data relevant to a particular task is usually sparse across the archive (e.g., only certain images are needed for the training of image representations, and only certain plaintexts are needed for argument mining), resulting in a variable data flow after preprocessing. To optimize GPU usage, a constant flow of data is required.
WARC-DL solves both problems: After loading web archive data from the memory cluster into the CPU~cluster for preprocessing, it streams the useful data into the GPU~cluster for interleaved processing.

\phantom{Pipeline}

\begin{figure}[t]
\begin{minipage}{\textwidth}
\vspace{31.5ex}
\end{minipage}
\end{figure}

\bslabel{Web Archive Processing Pipeline}
Figure~\ref{fig:pipeline} provides an overview of the six-step pipeline implemented by WARC-DL:
(1)~WARC files%
\footnote{\url{https://www.iso.org/standard/68004.html}}
are distributed to the CPU workers using PySpark.
(2)~FastWARC~\cite{bevendorff:2021}%
\footnote{\url{https://github.com/chatnoir-eu/chatnoir-resiliparse}}
is used to decompress and iterate the records. The records can optionally be filtered and CPU-bound preprocessing like feature extraction or tokenization can be performed.
(3)~Pickled record streams are sent to the GPU~Cluster via TCP.
(4)~The streams are converted into TensorFlow datasets and interleaved.
(5)~One option is to use a pre-trained Keras model to batch process the samples. The results are filtered and stored, including the original data extracted from the WARC datasets.
(6)~Alternatively, the samples can be used to train a Keras model after an optional filtering step, e.g., for duplicate removal.
The filtering in (2) and (6) implements a basic MapReduce.

The preprocessing steps and the pre-trained models used are fully customizable (an extension to frameworks other than TensorFlow is planned). The pipeline enables extraction of data from multiple modalities, including text for language models and images for computer vision models. It also supports the simultaneous extraction of linked web archive records, such as the text of a web page and the image that was originally linked on the page. This should enable efficient multimodal learning, where the pipeline transparently solves the problem of matching an image with its associated text. To optimize the ratio of allocated CPU to GPU resources depending on the model, a profiling method based on the TensorFlow Profiler is provided.

\printbibliography

@InProceedings{yang:2019,
  author =              {Hsiu{-}Wei Yang and Linqing Liu and Ian Milligan and Nick Ruest and Jimmy Lin},
  bibsource =           {dblp computer science bibliography, https://dblp.org},
  booktitle =           {JCDL 2019},
  editor =              {Maria Bonn and Dan Wu and J. Stephen Downie and Alain Martaus},
  publisher =           {{IEEE}},
  title =               {{Scalable content-based analysis of images in web archives with tensorflow and the archives unleashed toolkit}},
}

@InProceedings{bevendorff:2021,
  author =              {Janek Bevendorff and Martin Potthast and Benno Stein},
  booktitle =           {OSSYM 2021},
  editor =              {Andreas Wagner and Christian Guetl and Michael Granitzer and Stefan Voigt},
  month =               oct,
  publisher =           {International Open Search Symposium},
  site =                {CERN, Geneva, Switzerland},
  title =               {{Fastwarc: Optimizing large-scale web archive analytics}},
}

\end{document}